\documentclass[prl,a4paper]{revtex4}

\usepackage{amsmath,amssymb}
\usepackage{epsfig}

\newcommand{\numero}[1]{\noindent {\bf #1.}~}
\newcommand{\eq}{\begin{equation}}
\newcommand{\eqx}{\end{equation}}
\newcommand{\eqn}{\begin{eqnarray}}
\newcommand{\eqnx}{\end{eqnarray}}
\newcommand{\f}[2]{\frac{#1}{#2}}

\newcommand{\cor}[1]{\left\langle{#1}\right\rangle}

\newcommand{\DD}{{\cal D}}
\newcommand{\si}{\sigma}

\newcommand{\tr}{\mbox{\rm tr }}
\newcommand{\dery}{\f{\partial}{\partial Y}}
\newcommand{\al}{\alpha}
\newcommand{\bt}{\beta}

\begin{document}
\title{QCD Saturation Equations including Dipole-Dipole Correlations}

\author{Romuald A. Janik%
\footnote{e-mail: {\tt ufrjanik@if.uj.edu.pl}}
}
\affiliation{Institute of Physics, Jagellonian University, Reymonta 4,
  30-059 Krakow, Poland.}  
\author{R. Peschanski%
\footnote{e-mail: {\tt pesch@spht.saclay.cea.fr}}
}
\affiliation{Service de physique th{\'e}orique, CEA/Saclay,
  91191 Gif-sur-Yvette cedex, France\footnote{%
URA 2306, unit\'e de recherche associ\'ee au CNRS.}}

\begin{abstract}
We derive  two coupled non-linear evolution equations corresponding to the 
truncation of the Balitsky infinite hierarchy of saturation equations
after inclusion  
of   dipole-dipole correlations, i.e. one step beyond the
Balitsky-Kovchegov (BK)  
equation. We exhibit an exact solution for  maximal correlation which
still satisfies  the same 
asymptotic geometric scaling as BK but with the S-matrix going to 1/2
(instead of 0) in the full saturation region.
\end{abstract}

\maketitle

\numero{1}
In perturbative QCD, parton saturation, i.e. the modification of the
distribution of  
quarks and gluon distributions in a target, is known to lead to an
infinite set of  
coupled evolution equations in energy for the correlation functions of
multiple Wilson  
lines
\cite{Balitsky:2001gj}. This set of equations is expected to be
equivalent to the  
Jalilian-Marian, Iancu, McLerran, Weigert, Leonidov and Kovner
(JIMWLK)  functional  
equation \cite{JIMWLK}. In the approximation where correlation
functions  for  more  
than two Wilson lines factorize, i.e. $\cor{\tr U^2 \tr U^2}\sim
\cor{\tr U^2} \cor{\tr U^2}$, the  
problem reduces to an unique non-linear Balitsky-Kovchegov (BK) equation 
\cite{Balitsky:2001gj,Kovchegov} for the dipole density. This amounts
to considering a  
large-$N_c $ limit of  independent dipole-target collisions, e.g. on a
large nucleus.  
Recently it was shown that  the translation-invariant (no impact parameter 
$b$-dependence) non-linear BK equation lies in the
universality class  
of the Fisher and Kolmogorov-Petrovsky-Piscounov (F-KPP) equation
\cite{Munier:2003vc},  
leading to asymptotic traveling wave solutions in the transition
region to saturation. This provides a mathematical realization of the 
phenomenologically motivated \cite{Stasto:2000er} geometrical scaling. In 
addition, the BK equation leads to an  
 S-matrix element which  goes  to 0 in the full saturation
 region. Note that, if  one assumes that translation invariance is a
 good approximation for scattering near $b =0,$ this amounts to the
 black disk limit $\si_{el}/\si_{tot} (b\sim 0) \to 1/2.$

The problem we want to address is the study of QCD
saturation including the effects of nontrivial two-dipole
correlations. To this end we have to proceed one step further in the
hierarchy of Balitsky equations \cite{Balitsky:2001gj}. Our main
observation is that one can perform a truncation of this set at the
level of two-dipole correlations, as follows:
\begin{itemize}
\item We keep only dipole-like terms (i.e. we neglect higher
  multipoint traces like $\cor{\tr U^4}$ and $\cor{\tr U^6}$).
\item We keep full two-dipole correlations and only neglect independent three-
  and higher dipole correlations.
\end{itemize}
In this paper we derive the general  closed set of two non-linear equations
for the dipole  
density and the dipole-dipole correlation function, see equations
(\ref{e.ei},\ref{e.eii}) below. 
Moreover, using the 
translational invariance hypothesis (no  
$b$-dependence),  we find a particularly convenient form of these equations in
a Fourier  
transform representation see equations (\ref{simple},\ref{simple'}). We then 
exhibit a  
solution  in terms of a modified BK equation  when  
assuming {\it maximal} correlation, i.e. when the correlation stays
independent of the  
separation distance between the two dipoles see
(\ref{tspace1},\ref{tspace2}). We 
analyze the solutions, examine the modifications with respect to BK
ones and discuss the prospects of our study. 
As an aside we note that
some hierarchies of evolution equations have been studied in a
statistical physics context in \cite{PARISI}.

%%%%%%%%%%%%%%%%%%%%%%%%%%%%%%%%%%%%%%%%%%%%%%%%%%%%%%%%%%

\medskip

\numero{2}
Let us briefly recall the standard derivation of the BK equation in the
Balitsky's framework \cite{Balitsky:2001gj}.
One defines the dipole operator
\eq
\label{dipole}
\DD_{ij}=\tr U_{x_i} U^\dagger_{x_j}\ ,
\eqx 
where $x_{i,j}$ are   the transverse coordinates of
the end point quark and antiquark of a QCD  dipole. One 
calculates its evolution with rapidity $\partial/\partial Y$ using
contractions between pairs of $U$'s and virtual corrections for each
$U$ (see e.g. eqs. (119,120) in \cite{Balitsky:2001gj}.)

Within the infinite set of saturation equations,  the one for the
evolution of the dipole operator \cite {foot1} expectation value with
rapidity reads: 
\eq
\label{e.kovd}
\dery \cor{\DD_{01}} = \f{g^2}{8 \pi^3} \int d^2x_2 
\left[\cor{\DD_{02} \DD_{21}} - N_c \cor{\DD_{01}} \right] K_{021}
\eqx
where the kernel $K_{021}$ is
\eq
K_{021}= \f{x_{01}^2}{x_{02}^2 x_{21}^2}\ ,
\eqx
i.e. the real dipole splitting  part in the BFKL kernel
\cite{Lipatov:1976zz}. Let us   
now decompose the expectation values into 1-point functions and
connected correlation functions through 
\eqn
\cor{\DD_{02}} &=& d_{02} \\
\cor{\DD_{02} \DD_{21}} - \cor{\DD_{02}} \cor{\DD_{21}} &=& d_{0221}
\eqnx
The BK equation is  obtained by {\em neglecting} $d_{0221},$ i.e. the
connected part of the two-dipole correlation function, with two
coincident points at $x_2$.  

Our first goal is to keep now this term, and  derive an equation for
this quantity. In order to obtain a well-defined {\em closed} set of
equations, it 
turns out to be necessary to write an equation for
the correlation of arbitrary two dipoles, not constrained to have one
point in common. We thus consider more generally 
$\partial \cor{\DD_{02} \DD_{2'1}}/\partial Y$ and perform all
contractions between the unitary matrices. 
%The result is
%necessary to consider a slight generalization of the expression given
%for $2=2'$ in \cite{Balitsky:2001gj}. 

For the truncation, we proceed as follows. All single (1-point)
and double contractions within the same dipole give
essentially two copies of (\ref{e.kovd}) with the other dipole
included as a spectator. To this we have to add contractions between $U$'s
belonging to different dipoles. These  generate only non-dipole
terms of the form $\cor{\tr U^6}$ and $\cor{\tr U^4}$ which we neglect
by assumption.    
The resulting equation reads
\eq
\dery \cor{\DD_{02}\DD_{2'1}} =\f{g^2}{8\pi^3} \int d^2 x_3 \left[
  \cor{\DD_{03} \DD_{32} \DD_{2'1}} -N_c \cor{\DD_{02}\DD_{2'1}}
  \right] K_{032}+  \left[
  \cor{\DD_{2'3} \DD_{31} \DD_{02}} -N_c \cor{\DD_{02}\DD_{2'1}}
  \right] K_{132'} \ .
\eqx
When $2=2'$ this reduces to the expression in \cite{Balitsky:2001gj}
up to the neglected $\cor{\tr U^6}$ and $\cor{\tr U^4}$ terms.
Keeping track of all two-dipole correlations,  the expectation values
are evaluated giving 
\eq
\cor{\DD_{03} \DD_{32} \DD_{2'1}}=d_{03} d_{32} d_{2'1}
  +d_{0332}d_{2'1} + d_{322'1} d_{03} +d_{032'1} d_{32}\ .
\eqx
It is important to notice the appearance here of  two-dipole
correlation functions with  
four non coinciding points even if we would start with $2=2',$ due to the 
contractions between the first and third $\DD$ operators. This
justifies the need to get the corresponding 4-point evolution
equations. Indeed,   neglecting this term
would lead to spurious divergences in the equation.  
In this way we get: 
\eq
\dery d_{022'1}= \f{g^2}{8\pi^3} \int d^2 x_3 \left\{d_{03} d_{322'1}
  +d_{32}d_{032'1}- N_c d_{022'1} \right\} K_{032}  +\left\{ (0,1) \iff (2',2) 
\right\}
\eqx
Going from the Wilson line correlation traces $d$ to  dipole densities
  ($d_{01}=N_c(1-N_{01})$ and $d_{022'1} =N_c^2\ N_{022'1}$) we get the
  final closed set of equations:
\eqn
\label{e.ei}
\!\!\!\!\!\!\dery N_{022'1}&=& \f{g^2 N_c}{8\pi^3} \int d^2 x_3 
\left\{\left[N_{322'1}
  \!+ \!N_{032'1} \!- \! N_{022'1} \right] K_{032}  - \left[ N_{03}N_{322'1}  
\!+ 
\! 
N_{32}N_{032'1} \right] K_{032} \right\} + 
\left\{  (0,1) \iff (2',2) \right\}  \\
\label{e.eii}
\dery N_{01} &=& \f{g^2 N_c}{8\pi^3} \int d^2 x_2 \left[N_{02} +N_{21}
  -N_{01} -N_{02}N_{21} -N_{0221} \right] K_{021}
\ .
\eqnx

\medskip

\numero{3}It is well known \cite{Kovchegov} that after imposing translational 
invariance (i.e. restricting to impact-parameter independent regime) and
going over to momentum space the BK equation greatly simplifies. We
will perform an analogous procedure for our set of equations
(\ref{e.ei})-(\ref{e.eii}). Let  us introduce the following
2-dimensional Fourier transfoms
\eqn
N_{01}%(x_0,x_1)
 &=& 
x_{01}^2 \int \f{d^2 k}{2\pi}\ e^{ik\cdot x_{01}}\ N_k \\
N_{022'1}%(x_0,x_2,x_{2'},x_1)
 &=&
 x_{02}^2 x_{2'1}^2 \int  \f{d^2
  q}{2\pi} \f{d^2 q'}{2\pi} {d^2 Q}\
e^{iq\cdot x_{02}}\ e^{iq'\cdot x_{2'1}}\ e^{iQ\cdot
  (x_{02'}+x_{21})}\ N_{qq',Q} 
\eqnx
Here $q$ and $q'$ are 2-vectors conjugate to the dipole vectors,
while $Q$ is the variable dual to the $2$-vector drawn between the center of
masses of the two dipoles.   
The nonlinear terms now become local in momentum space, while the
linear terms reduce essentially to the BFKL kernel \cite{foot2}. 
We get finally:
\eqn
\label{simple}
\dery N_q &=&  \f{g^2 N_c}{4\pi^2}\ \{2\chi(-\partial_L) N_q -N_q^2
- \int d^2 Q\ N_{(q\!-\!Q) (q\!-\!Q),Q}\} \\
\label{simple'}
\dery N_{qq',Q} &=& \f{g^2 N_c}{4\pi^2}\ \left\{2\chi(-\partial_L)  
N_{qq',Q}+2\chi(-\partial_{L'})  N_{qq',Q}
- (N_{q+Q}+N_{q-Q}+N_{q'+Q}+N_{q'-Q}) N_{qq',Q}\right\}\ ,
\eqnx
where $L \equiv \log (q),$ 
\begin{equation}
\chi(\gamma)=2\psi(1)-\psi(\gamma)-\psi(1-\gamma)
\label{chi}
\end{equation}
and $\chi\left(-\partial_L\right)$ is an integro-differential
operator which may be defined \cite{Munier:2003vc} 
with the help of a formal series expansion around some given
$\gamma_0$ between 0 and 1, {\it i.e.} for the principal branch
of the function $\chi$. Note that
the integral term in the first equation (\ref{simple}) corresponds to
the integration of a dipole-dipole correlator (5) with the kernel
$K_{021}$ over the separation distance of the two dipoles with 
one coinciding endpoint.

The initial conditions for  correlations can be formulated in the
variable $Q$ by assuming a gaussian form $e^{-l_{corr.}^2Q^2}$
corresponding to a transverse correlation length  $l_{corr.}.$ 
In order to illustrate the impact of
these correlations on the problem, let us  consider  the limiting case
of {\it maximal correlations} i.e. when  
the dipole-dipole  correlations are local in~$Q$ ($l_{corr.} \to \infty$):
\eq
N_{qq',Q} = N_{qq'} \delta^2(Q)\ .
\eqx
Then the equations simplify and we get
\eqn
\label{e.iext}
\dery N_q &=&  \f{g^2 N_c}{4\pi^2}\left\{2\chi(-\partial_L) N_q -N_q^2
- N_{qq}\right\}  
\\
\label{e.iiext}
\dery N_{qq'} &=&  \f{g^2 N_c}{4\pi^2}\left\{2\chi(-\partial_L)  
N_{qq'}+2\chi(-\partial_{L'})
N_{qq'} -2(N_q+N_{q'}) N_{qq'}\right\}\ .
\eqnx
Remarkably enough this set of equations admits an exact solution (apart from  the 
trivial solution $N_{qq'} = 0$) using the
ansatz $N_{qq'}=\lambda N_q N_{q'}$. Note that this form may be quite
plausible physically as typically these dipoles are separated by very large
distances, therefore we do not expect strong dependence of the
correlations on the {\em relative} sizes.
Plugging the ansatz into equations (\ref{e.iext})-(\ref{e.iiext}) we
obtain $\lambda=1$ and find the following system
\eqn
\label{sol2}
N_{qq'} &=& N_q  N_{q'} \\
\label{sol1}
\dery N_q &=&  \f{g^2 N_c}{4\pi^2}\left\{2\chi(-\partial_L) N_q -{\bf
  2}N_q^2\right\} \ ,
\eqnx
or equivalently in position space:
\eqn
\label{tspace1}
N_{022'1}&=& N_{02}  N_{2'1} \\
\label{tspace2}
\dery N_{01} &=& \f{g^2 N_c}{8\pi^3} \int d^2 x_2 \left[N_{02} +N_{21}
  -N_{01} - {\bf 2}  N_{02}N_{21}\right] K_{021} \ .
\eqnx
In fact this solution can be extended to the full $b$-dependent set of equations 
(\ref{e.ei},\ref{e.eii}).

It is important to note the factor ${\bf 2}$ in front of the
non-linear term which is the  consequence of the
correlations. Note that a modification of the BK equation of
the type (\ref{tspace2}) but with ${\bf 2}$ replaced by some constant
has been proposed in \cite{LEVIN} on phenomenological grounds to
account for the effect of correlations in nuclei. Our approach gives a
derivation of such a modification from the JIMWLK-Balitsky framework
in the special limit of infinite range correlations and opens up a way to
study corrections due to finite correlation length. 

The solution of the equation (\ref{tspace2}) can be
obtained from the solution of the {\em uncorrelated} BK equation
through the relation
\eq
\label{BK}
N_{01}(l_{corr.} \to \infty) \equiv \f{1}{2}N_{01}^{BK} \ . 
\eqx
 The $S$
matrix is consequently $S=1-N=1-\f{1}{2}N^{BK}$ and satisfies the
equation:
\eq
\label{smatrix}
\dery S_{01} = \f{g^2 N_c}{8\pi^3} \int d^2 x_2 \left[ { 2} 
(S_{02}- {\scriptstyle {\f12}}) \times
 (S_{21}- {\scriptstyle {\f12}}) - (S_{01}-
  {\scriptstyle{\f12}})\right] K_{021} \ . 
\eqx

A consequence of the above equations is that the solution
for the case with 
maximal correlations leads to a saturation regime where the S-matrix goes to 
$\scriptstyle{\f12}$ instead of $0.$  Physically, it means that the
ratio of the elastic over total cross-section (at small impact
parameter at least) $\si_{el}/\si_{tot}
 (b\sim 0) \to {\scriptstyle{\f14}}$  instead of
$\scriptstyle{\f12}$ for the BK equation. It is remarkable that the fixed point 
solution $S \to {\scriptstyle \f 12}$ saturates the so-called Pumplin bound 
\cite{pumplin}, which states that $\si_{el} + \si_{dd} \le 
{\scriptstyle{\f12}}\si_{tot},$ where $\si_{dd}$ is the contribution of inelastic 
diffractive channels to the total cross-section. Indeed, a simple calculation 
\cite{pumplin} gives  $\si_{el} =  \si_{dd} = 
\scriptstyle{\f14}\si_{tot}={\scriptstyle{\f14}}.$ It is interesting to notice 
that 
the phenomenological  extraction of  dipole-proton S-matrix values 
\cite{Munier:2001nr}
seems to be consistent with $S \ge \scriptstyle{\f12}$.

Another interesting aspect is  that 
geometrical
scaling and more generally, the transition to saturation is {\it not}
expected to be modified, since it follows directly from the exact relation 
(\ref{BK}). 
It is yet another consequence of the universality properties of the 
traveling wave solutions \cite{Munier:2003vc}.
The expectations from the
mathematical properties of  non-linear equations  discussed in
\cite{Munier:2003vc} show that the behaviour of  the solution in that
region are universal, i.e. independent of the precise form of the non-linear
damping. 
Indeed, since equations
(\ref{simple})-(\ref{simple'}) have a similar  linear term as in the BK
equation and lead to geometrical scaling both in the maximal
correlation regime and in the regime of no correlations at all,
therefore it is tempting to conjecture that geometrical scaling holds
even when the assumption of maximal correlations is relaxed.
This point 
certainly 
deserves further study.

\medskip

\numero{4} 
In conclusion, we have shown that a consistent truncation can be performed on 
the
infinite hierarchy \cite{Balitsky:2001gj} of Balitsky equations for
$n$-point  correlators of Wilson lines by keeping track of nontrivial
two-dipole correlations. It results in a system of two coupled non-linear
equations. It is interesting to note that the assumption of maximal
correlations does not modify the traveling wave picture of
geometrical scaling.  
%transition near  the saturation scale
%which is not modified by correlations. 
In contrast, the fully
saturated region is strongly modified, since the S-matrix admits a
different limit. 

Recently, it has been argued that a certain treatment of fluctuations
beyond the BK equation may lead to deep modifications including strong
geometrical scaling violations \cite{shoshi}. We do not find this
phenomenon when including  dipole-dipole correlators with maximal
(very long-range) correlations. It would be  interesting to look
what is the behaviour of the solution for finite range correlations.
A numerical simulation of  our system of equations
(\ref{simple})-(\ref{simple'}) seems feasible and could be helpful. 

An important point to study is the dependence of the solutions on the initial 
conditions, in particular the stability of the fixed point solution
for maximal  
correlations when one starts from generic initial conditions (apart from the 
initial condition with no correlations which trivially reduces to the
standard BK  
equation). Interestingly enough it seems not out of reach to generalize the 
maximal correlation ansatz (\ref{tspace1}) 
to multipoint
correlations in the  
JIMWLK context.

It is interesting to ask the question whether the connected dipole
correlation functions could be directly related to physical
observables. For instance, Fourier transforming back the function
$N_{qq',Q}$ (see (\ref{simple'})) over $q,q'$  leads to a  relation with the 
interaction amplitude of 2 dipoles with the target. By crossing symmetry 
\cite{Mueller:fa}, some information can be obtained on    the semi-inclusive
$dipole(r) \to dipole (r') +X$ scattering   with transverse momentum $Q$ on the 
target. This problem
deserves more study, since it could give an interesting link with inelastic
diffractive processes.  

\medskip

\noindent{\bf Acknowledgements.} We thank Edmond
Iancu for informing us of ref. \cite{LEVIN} and St\'ephane Munier for
ref. \cite{PARISI}. RJ would like to thank Service de
Physique Th{\'e}orique, Saclay
for hospitality when this work was carried out. RJ was supported in
part by the KBN grant 2P03B08225 (2003-2006) and by the ECONET programme No 
08155PC of the French Ministry of Foreign Affairs.


\begin{thebibliography}{10}
\bibitem{foot1}
Here we use the ``S-matrix'' version of the equation. We pass to
the  ``T-matrix'' form below.
\bibitem{foot2}
In the
  case of correlations this can be checked by noting that a convenient basis is 
provided by bilinears of the BFKL  $SL(2,{\mathbb C})$  eigenfunctions 
\cite{Lipatov:1985uk}: 
  $N_{022'1}=\sum n^{hh'}(z_\al,z_\bt) E^h(z_{0\al},z_{2\al})
  E^{h'}(z_{2'\bt} z_{1\bt})$, acting with the linear kernel and only
  then going over to momentum representation.
\end{thebibliography}
\end{document}